\documentclass[aps,pre,twocolumn,floatfix,showpacs,showkeys]{revtex4}
\usepackage{amsmath,amsfonts,amssymb,mathrsfs,graphicx}
\begin{document}
\title{Entangled Quantum Networks}

\author{Fariel Shafee}
\affiliation{ Department of Physics\\ Princeton University\\
Princeton, NJ 08540\\ USA.} \email{fshafee@princeton.edu }

\begin {abstract}
We present a fairly simple method of obtaining a completely
entangled lattice of qubits, which are relevant in quantum computing
and may be important also in other uses of quantum neural networks,
using a modified form of the controlled-NOT quantum gates connecting
nearest neighbors for computational economy. A normal c-NOT gate,
though unitary and having the simplicity of a single operation of
flipping the controlled qubit, is time consuming, as it produces
complex mixing coefficients. Therefore a slightly modified form of
the gate, which is named $c^\prime-NOT$ gate here, is used, which
inverts the phase as well as the state of the controlled qubit when
the controlling qubit is excited.  It too gives a manifestly unitary
transition matrix for each updating of the network while keeping all
the numbers produced in the operations real. The dynamics leads to a
completely entangled state of the qubits in the system with variable
coefficients for the superposition of the states of the qubit nodes
in different circumstances. Simulation shows a surprising property
of the dynamics of the network, viz. the possibility of obtaining
the initial state by a method of back-projecting the complicated
entangled states that evolve after thousands of updating of the
entire network involving modifications of each qubit through
interactions with the neighbors through the quantum gates. We also
prove that it is not possible for a sequence of unitary operators
working on a net to make it move from an aperiodic regime to a
periodic one, unlike some classical cases where phase-locking
happens in course of evolution. However, we show that it is possible
to introduce by hand periodic orbits to sets of initial states,
which may be useful in forming dynamic pattern recognition systems.

\end{abstract}

 \vspace*{1cm}

\maketitle

\section{INTRODUCTION}

    Quantum computing promises \cite{NC1} to make a breakthrough in the capacity of computers
    to hold numbers and in the speed of processing on account of the entanglement of states.
    As neural networks can be conceived as dedicated hardware which can also process
    information for pattern recognition and other AI tasks,
    including associating specific memories with its internal
    dynamism, for later recovery or usage, quantum neural networks are also
    candidates for exhaustive study. However, this problem of transition to quantum
    neural nets may best be approached in stages, on account of the complications that
    different attributes of a quantized system bring about. The evolution of a
    semiclassical neural network model with an action-potential-like quantum interaction
    that triggers classical nodes in a stochastic manner have previously been simulated
    and the results agree in part with classical integrate-and-fire models while
    gaining some new features from the quantum characteristics of the system \cite{FS1}.
    A quantum neural network with nearest neighbor nodes connected by c-NOT gates,
    as might be expected to be present in a quantum computer, have also been
    studied \cite{FS2}. However, in both these investigations complete entanglement
    of all the nodes of the system have not been attempted; each qubit has been
    measured independent of the others, though in the latter work the interaction
    proceeded by unitary quantum c-NOT gates between nearest neighbor nodes. The neglect of
    entanglement of the whole lattice was unavoidable in these case-studies,
    as the concentration was on the origin and pattern of periodicity of the
    dynamics of the networks, which required a fairly large number of nodes,
    and it is virtually impossible to do simulation experiments with the
    corresponding huge number of completely entangled states of the system.
    We now present some results with a more complex model of a completely
    entangled quantum network on a computable scale. The simplest nontrivial
    net with a small number of nodes and symmetrical boundary conditions is
    constructed and results of computer simulations are found. It is seen
    that with normal c-NOT gates we need considerably more computing time
    than with a modified $c^\prime-NOT$ form of the gates. We also show why a quantum
    system in general cannot develop any dynamic periodicity which is often
    seen in classical models achieving phase-locking of the nodes after passing
    through an aperiodic regime. We then show how {\em ab initio} periodicity may be
    introduced by hand.

\section{ AN ENTANGLED QUANTUM NETWORK  MODEL}

    We take a $nXn$ lattice with usual periodic boundary conditions, so that it can
    effectively mimic a bigger lattice in some respects. So we have $N= nXn$ independent
    nodes, each a quantum qubit, e.g. a spin-1/2 object. We connect each node to
    its neighbor through c-NOT gates. We are adopting this simple scheme to see
    how such a system with no pre-design for any specific purpose would behave
    under different inputs. In other words, we want to investigate the robustness
    of the input in a symmetric quantum system that treats all inputs equally and
    also allows maximal entanglement regardless of input.

    The state of the system at any instant can be represented by

\begin{equation} \label{eq1}
|\psi \rangle = \sum_I a_I |\psi_I \rangle
\end{equation}

    where the complete set of unentangled product basis states    includes all
    the possible combinations of the type

\begin{equation} \label{eq2}
|q_1q_2 ....q_N \rangle
\end{equation}

    where $N = n^2$

    and each of the qubits can be either in state $|1\rangle$ or in state $|0\rangle$.

    So, initially we may have a pure state with only one $a_I   = 1$,  and all
    others zero, but as the entanglement proceeds through the c-NOT gates between
    the nodes, we expect that all or a subclass of states may become entangled.
    We can of course also choose an entangled initial state, by choosing a
    non-factorizable superposition of the product states.

    We have to consider the effect of each gate at each time step on every $a_I$.
    The c-NOT gate flips the controlled qubit if the controlling qubit is
    in state $|1\rangle$, while it does nothing if the controller is in state $|0\rangle$.
    The controller node is unchanged.

    Hence, during simulation we can take each node in turn and consider the effect
    on all $a_I$ as the neighbors of this node act on this node. The procedure is
    similar to that described in the earlier unentangled version.

    As in the previous case, we first do the flipping in a continuous manner by
    choosing for each time step   dt the transition submatrix for a small change

\begin{eqnarray} \label{eq3}
     A=
\left(
\begin{array}{cc}
1  &   0  \\
0  &   \exp(i \epsilon  \sigma_1  dt)
\end{array}
\right) =
     \left[
     \begin{array}{cccc}
     1  &  0  & 0  &  0 \\
     0  &  1  & 0  &  0 \\
     0  &  0  & 1  &  i \epsilon dt  \\
     0  &  0  & i\epsilon dt & 1
     \end{array}
     \right]
\end{eqnarray}

    However, in the previous work, because the nodes remained unentangled and the
    only effect of the gate was simply to flip each node independently, the state
    space consisted of only $2 n^2$ components (i.e. polynomial in $n$), in contrast to
    the exponential number in the entangled case. It is not possible to handle a
    large lattice like $40X40$ in this case. In this work we have considered the
    smallest nontrivial lattice, i.e. a $3X3$ lattice.  As we have remarked earlier,
    because of periodic boundary conditions, this is in some limited respects
    equivalent to an infinite lattice.
    Even for a $4X4$ lattice, there are now $216$ product states to upgrade at each step,
    which is not computationally economic with a classical computer

    To optimize computing time we have first linearized the $2-d$ label of each
    qubit $( i , j)$ to a single $1-d$ label by choosing a sequence, and then we have
    constructed our label $I$ (stated above) by simply taking the sum:

    \begin{equation} \label{eq4}
I = \sum_i 2^i
\end{equation}

    where the sum is over only those qubits for which the state is $|1\rangle$ and $i$ is the
    linear sequential position label of the qubit (0 to N). This permits ascertaining
    the state of any qubit in a particular position $i$ with a single bitwise AND  (\&)
    operation and speeds up the simulation process. This also permits putting
    on any initial state, pure or entangled, by choosing the right combination of $I$'s.

\section{PERIODIC AND APERIODIC REGIMES}
 Before we do our simulations, let us point out a behavior we can anticipate from
 purely theoretical considerations for any quantum net on which a particular
 sequence of unitary operators work repeatedly.

 {\bf Lemma}:     It is not possible by a repeated sequence of unitary operators
 to move any system from an aperiodic regime to a periodic one.

 {\em {\bf Proof}}:  The product of any given sequence unitary operators $U_i$  is equivalent
 to a single unitary operator, say $U$.

 Let  $| i \rangle$  be a vector in the orbit of $U$  in the periodic regime. Now if we
 operate on $| i \rangle$ by

 \begin{eqnarray}\label{eq5}
 U^\dag  = U^{-1} \\
 U^{-1} |i \rangle = |j\rangle  \nonumber
 \end{eqnarray}

 where  $|j \rangle$ must be on the orbit.  On the other hand if the state was reached
 from the aperiodic regime, then we must also have, with the same inverse
 operation, a transition to a state in the aperiodic regime. This is not
 possible, because $U$  and its inverse are both linear operators and must
 give unique results whichever state they operate on.

 Hence we would not expect any transition to a periodic system in our simulation.

 \section{SIMULATION RESULTS AND A MODIFIED GATE}

 For a $3X3$ lattice there is only one interior point. So if we initially
 excite the whole periphery we have initially

  \begin{equation}\label{eq6}
  a_{495} = 1.0
  \end{equation}

 As
\begin{equation}\label{eq7}
 495 = 111101111_2
 \end{equation}

 To ascertain the importance of unitarity in these simulations we first
 deliberately used a nonunitary series of operations with $\epsilon$ imaginary. Our
 simulations show that even if we begin with a pure state, the nodes get
 entangled quickly after only a few steps, but after a sufficiently long time
 the system degenerates to a uniform state with all $a_i= 1/\surd N$.

 This seems to be because the operator {\bf A} above, which tries to form a
 continuous c-NOT gate in place of the unitary discrete one:

\begin{equation}\label{eq8}
 C=
 \left(
 \begin{array}{cc}
 1 & 0 \\
 0 & \sigma_1
 \end{array}
 \right)
 \end{equation}

 using the imaginary parameter $\epsilon$,  becomes a nonunitary one. Hence, unlike
 a unitary c-NOT operator,  this discrete version with an imaginary parameter
 does not have all eigenvalues of modulus 1. What is happening here is the
 emergence of the eigenstate corresponding to the highest eigenvalue, as is
 usually the case for multiple operations of a nonunitary operator. However,
 since the full matrix even for the $3X3$ net must be $512X512$, we cannot check
 it computationally or analytically. One can argue from symmetry that the highest
 eigenstate must be the symmetric one, though we are not aware of any mathematical
 theorem that justifies this hypothesis.

 To keep the computational expenses minimal, we shall adhere to real matrices.
 We next try to construct the infinitesimal form of a unitary matrix representation
 with a $c^\prime-NOT$ gate which is defined as a quantum gate that reverses the phases of
 the flipped infinitesimal changed coefficients:

\begin{eqnarray} \label{eq9}
U= \exp(i H dt) = 1 + i \epsilon dt \left(
\begin{array}{cc}
0  &   0  \\
0  &  \sigma_2
\end{array}
\right) = \left(
\begin{array}{cccc}
1  &   0  &  0  &  0 \\
0  &   1  &  0  &  0 \\
0  &   0  &  1  &  \epsilon dt \\
0  &   0  &  -\epsilon dt   &  1
\end{array}
\right)
\end{eqnarray}

 where $H$ is a hermitian Hamiltonian, making $U$ unitary.

 We first try some pure initial states and note the biggest components
 after $1000$ time loops.

\begin{equation} \label{eq10}
|27 \rangle \rightarrow 0.71 (|27 \rangle) + 0.21 ( |11\rangle + |19
\rangle +|25 \rangle + |26 \rangle)
\end{equation}

where we have omitted the smaller order terms.

 We notice immediately that the initial state has retained its dominance even
 after $1000$ time steps, and also that the next to leading states are separated
 from it by just a single $1$ in a neighboring qubit.

 For another single state $|17\rangle = |1+16\rangle$ i.e. the one with only a corner and
 the middle qubit of the lattice initially excited, the final state is
 (highest amplitude terms):

 \begin{equation} \label{eq11}
|17 \rangle \rightarrow 0.27 (|27 \rangle) + 0.23 ( |19\rangle +|25
\rangle)
\end{equation}

 We note that this time the original state has disappeared from the list of
 dominant states finally, but the terms now dominating do not have a clear choice,
 and with the AND operation on the bits of the dominating ones we get back
 the initial $|17\rangle$  !

\begin{equation}\label{eq12}
 (10001)  = (11011)  \& (1011)  \& (1101)
 \end{equation}

\section{ CREATION AND DETECTION OF  ENTANGLED  INPUT STATES}

 In the simulation above we have assumed that the system acquired an entangled
 state as input. Rabitz et al \cite{SR1} have presented a method of obtaining superposition
 of states from a ground state in a molecular system. In a similar spirit we here
 indicate how it may be possible to get arbitrary combinations of states in our model
 in a general quantum network.
 Let us consider a matrix, which we call an "extended unitary matrix", as given below:
\begin{equation}\label{eq13}
U(R, x^\prime ) = \left(
\begin{array}{cc}
R  &   b |x^\prime \rangle \langle n+1| \\
0  &   |n+1\rangle \langle n+1|
\end{array}
\right)
\end{equation}

 with the normalization

\begin{equation}\label{eq14}
|a|^2 + |b|^2 =1
\end{equation}

 This operator matrix acts on the $(n+1)$-dimensional basis with the last vector
 $|n+1\rangle$ an auxiliary  vector not related to the $n$-dimensional entangled vector space.

 Then, given any state   $|x\rangle$, we get the normalized new state

 \begin{equation} \label{eq15}
 |x^{\prime \prime} \rangle = a R |x\rangle + b |x^\prime \rangle
 \end{equation}

 with $R$ a $nXn$  unitary operator restricted to giving a vector orthogonal to the
 new vector $|x^\prime \rangle$ to be superposed. Of course in the simplest cases we can choose
 $R$ to be just the unit operator, and the constants $a$ and $b$ can be chosen to be   .
 Obviously this generalized unitary operator in the extended $(n +1)$-dimensional
 vector space separately maintains the length of the $n$-dimensional vector of the
 active system and the single auxiliary vector, which may be an additional dummy
 component of the system.

 It is well-known that a complete set of gates, e.g. c-NOT gates, phase gates and
 Hadamard gates, can \cite{BA1}  simulate any unitary operator. One can trivially extend
 the arguments to produce our extended unitary operator with such gates too. So a
 physical realization is not, at least in theory, an insurmountable problem.

 The entangled states, despite being superpositions, are pure states. Hence,
 in principle, the detection of the entangled states is no more complicated than
 that of single states in the usual basis of the product basis of single spins,
 provided we rotate the basis to a one that has the chosen vector as a basis vector.
 One can then use Grover's search procedure \cite{GO1}  to detect the presence of any
 particular state. Alternatively, we can determine filtering matrices that perform
 the same operation directly in the original basis by adapting the sign-reversing
 and diffusion matrices of Grover for the superposed state taking the appropriate
 linear transformations. The detection process in quantum computation is of course
 only stochastic, as originally proposed by Deutsch \cite{DE1}.

\section{ CONCLUSIONS}

We see that a network of qubits which is allowed to get completely
entangled through an arbitrarily constructed simple nearest neighbor
interactions through quantum $c^\prime-NOT$ gates, does indeed do so
in general, with a smearing of the excitation from the initially
excited nodes to its neighbors, as expected. However, the memory of
the initial state seems to be preserved in nontrivial ways depending
on whether it is a pure initial state or an entangled one. The
method of back-projection to the initial state is fairly simple. It
would be interesting to see how entanglement with low noise can be
filtered out in this system. It might also be interesting to
investigate if a quantum net can serve as a filter separating
entangled and separable states, by criteria similar to or different
from those proposed recently by Doherty et al [3].

We have shown that unlike a classical network which may often
produce phase locking among its nodes and make a transition from an
aperiodic regime to a periodic one, a quantum system operated on
repeatedly by the same sequence of unitary operators cannot make
such a transition, but must always remain in the aperiodic or the
periodic regime.

We have indicated that periodic dynamic behavior may be injected
into the system at will by choosing the right operator, i.e. a
suitable unitary operator that rotates some or all states with time.
This may be achieved by choosing the appropriate connectivity among
the nodes, which remains to be studied in detail.  By giving certain
subclasses of the set of states an identical period, it may become
possible to use such a net for pattern identification over a huge
data base created by the entire set of separable and entangled
states. The advantage over a classical system remains the great
expansion of the basis in using entangled qubits in place of
independent classical binary registers.

The author would like to thank Todd Brun of the Institute for
Advanced Study for useful discussions.


\begin{thebibliography}{}
\bibitem{NC1} M.A. ~Nielsen and M. Chuang, {\em Quantum
computation and quantum information} (Cambridge U.P., NY, 2000)
\bibitem {FS1} ~Shafee, F., "Semiclassical Neural Network", arxiv.org,
quant-ph/0202015 (2002)
\bibitem{FS2} ~Shafee, F., "Neural Networks with c-NOT Gated Nodes",
arxiv.org, quant-ph/0202016 (2002)
\bibitem{DO1} ~Doherty, A.C. et al, "Distinguishing entangled and
separable states", arxiv.org,quant-ph/0112007 (2001)
\bibitem{SR1} ~Schirmer, S.G., H. Rabitz et al., "Quantum control using
sequence of simple control pulses", arxiv.org,
quant-ph/0105155(2001)
\bibitem{BA1} ~Barenco, A. et al. , Phys. Rev. A {\bf52}, 3457 (1995)
\bibitem{GO1} ~Grover, L.K., " A fast quantum-mechanical algorithm for
database search", {\em Proc. 28th Annul ACM Symposium on the Theory
of Computing (STOC'96)} p.212 (ACM, Philadelphia, 1996).
\bibitem{DE1} ~Deutsch, D., Proc. Royal Soc. Lond. A {\bf 400}, 97 (1985)

\end{thebibliography}
\end{document}